\documentstyle[12pt,aaspp4]{article}
\tighten

\lefthead{Band, Hartmann \& Schaefer}
\righthead{GRB No Host Problem II}

\def\etal{et al.}

\begin{document}

\title{A STATISTICAL TREATMENT OF THE GAMMA-RAY BURST ``NO HOST
GALAXY'' PROBLEM: \\ II. ENERGIES OF STANDARD CANDLE BURSTS} 
\author{David L. Band}
\affil{CASS 0424, University of California, San Diego, La Jolla, CA  
92093; dband@ucsd.edu}
\author{Dieter H. Hartmann}
\affil{Department of Physics and Astronomy, Kinard Laboratory of
Physics, Clemson University, Clemson, SC 29634;
hartmann@grb.phys.clemson.edu} 
\author{Bradley E. Schaefer}
\affil{Department of Physics, Yale University, New Haven, CT 06520; 
schaefer@grb2.physics.yale.edu}
\centerline{\it Received 1998 July 8; accepted 1998 November 6}
\centerline{To appear in the 1999 April 1 issue of}
\centerline{{\it The Astrophysical Journal} (volume {\bf 514})}
\begin{abstract}
With the discovery that the afterglows after some bursts are
coincident with faint galaxies, the search for host galaxies is no
longer a test of whether bursts are at cosmological 
distances, but rather a test of
particular cosmological models.  The methodology we developed to
investigate the original ``no host galaxy'' problem is equally valid
for testing different cosmological models, and is applicable to the
galaxies coincident with optical transients.  We apply this
methodology to a family of models where we vary the total energy of
standard candle bursts.  We find that total isotropic energies of
$E<2\times10^{52}$~erg are ruled out while $E\sim10^{53}$~erg is
favored. 
\end{abstract}
\section{INTRODUCTION}
The absence of the host galaxies expected under the simplest
``minimal'' cosmological gamma-ray burst model was first advanced as a
challenge to the cosmological hypothesis of burst origin (Schaefer
1992), but with the evidence from the recently-discovered optical
transients (OTs) that some, and probably all, bursts are at cosmological
distances,
the search for host galaxies is now a tool for learning where bursts
occur.  The minimal model assumed that bursts are standard candles
which did not evolve and that they occur in galaxies at a rate
proportional to the galaxy luminosity (e.g., Fenimore et al. 1993).
Because of a dispute as to whether there was indeed a ``no-host''
problem for the minimal model (Larson \& McLean 1997), we developed a
statistical methodology which compares the hypotheses that host
galaxies are or are not present (Band \& Hartmann 1998, henceforth
Paper~I).  This methodology clearly demonstrated the obvious point
that one can only test a well-defined model.  A preliminary
application of this methodology showed that the galaxies predicted by
the minimal cosmological model were indeed absent. 

As a result of the galaxies coincident with the OTs, and the
magnitudes and redshifts of these galaxies, there is little doubt that
some (and by Occam's Razor, probably all) bursts are cosmological but
the minimal cosmological model is clearly too simple. The methodology
we developed tests a particular cosmological model against the
hypothesis that the host galaxies predicted by this model are 
not present; this methodology can be
generalized to compare different models.  The methodology includes a
finite-sized ``error box'' for the particular burst under
investigation, which would seem to be inappropriate for bursts
followed by OTs whose positions are known exceedingly well. However,
the error box actually consists of the burst localization uncertainty
and the model-dependent region around the host galaxy 
in which the burst is expected to
occur. For example, some models may require the burst to occur at the
center of the host galaxy (e.g., a flare by an otherwise dormant AGN)
while other models may permit bursts to occur in an extended halo
surrounding the host galaxy. 

In this paper we make the simplest modification to the minimal model.
Bursts are still standard candles which occur in galaxies at a rate
proportional to the galaxy's luminosity, but we vary the intrinsic
brightness of the standard candle.  Such a model would be consistent
with the observed burst intensity distribution only if the source
density is allowed to evolve (Fenimore \& Bloom 1995).  Because of the
redshift associated with GRB~970508 (Metzger et al. 1997), the source
models in which the death of a massive (therefore short-lived) star
gives birth to the burst progenitor (e.g., a neutron star), and the
implications of the host galaxy issue, a model as been proposed where
the burst rate is proportional to the cosmic star formation rate
(Totani 1997; Wijers et al. 1998; Hartmann \& Band 1998; Krumholtz,
Thorsett \& Harrison 1998; Che, Yang \& Nemiroff 1998).  In these new
cosmological models, bursts occur at greater redshifts, and
consequently their intrinsic brightness must increase.  Here we
determine what intrinsic brightness is consistent with the host galaxy
observations.  Bursts are standard candles in the model we study,
which is clearly not the case, as shown by Table~1.  In Table~1 we
include GRB~980425, even though this burst, associated with a peculiar
supernova (Galama et al. 1998), is most likely from a population
different from most bursts.  In future studies we will include
luminosity functions in our analysis.  Nonetheless, the analysis here
demonstrates decisively that the average burst energy is much greater
than previously thought. 

Based on some of the same data we use here, Schaefer (1998) also
concludes that if bursts are in galaxies, then they must intrinsically
be two orders of magnitude brighter than predicted by the minimal
model. Schaefer calculates the fraction of the model-dependent host
galaxy distribution which is fainter than the brightest observed
galaxy; if only host galaxies are present, then the average of this
fraction should be 1/2 if the host galaxy model is correct. To 
compensate for the presence of unrelated
background galaxies, Schaefer weights this fraction for each burst 
based on the brightness ratio of the expected host and background 
galaxies. 
 
Since the statistical methodology is derived in Paper~1, here we only
review the basic formulae (\S 2.1). Because many of the cosmological
models push the host galaxies out to higher redshifts, we can no
longer rely on the Euclidean $r^{-2}$ law to relate the intrinsic and
observed galaxy brightnesses, but we must include both $k$- (spectrum
redshifting) and $e$-
(evolution) corrections; the sources of our astronomical data are
presented in (\S 2.2).  In \S 3 we analyze different datasets, and
discuss the results in \S 4. 
\section{METHODOLOGY}
\subsection{The Likelihood Ratio}
In Paper~I we presented a Bayesian odds ratio which compares the
hypothesis $H_{\rm hg}$ that both host galaxies of a specific
cosmological model and unrelated background galaxies are present in
burst error boxes to the hypothesis $H_{\rm bg}$ that only background
galaxies are present.  The odds ratio for a set of $N$ bursts
\begin{equation}
O_{\rm hg,bg} = {{p(H_{\rm hg})}\over{p(H_{\rm bg})}} \prod_{i=1}^N
   {{p(D_i \,|\, H_{\rm hg})}\over{p(D_i \,|\, H_{\rm bg})}}
\end{equation}
consists of two factors.  The first is the ratio $p(H_{\rm hg}) /
p(H_{\rm bg})$ of the ``priors,'' the probabilities that each
hypothesis is correct, evaluated before the new data were acquired.
The second is the ``Bayes'' factor $\prod p(D_i \,|\, H_{\rm hg}) /
p(D_i \,|\, H_{\rm bg})$, the ratio of the likelihoods for each
hypothesis. The expression $D_i$ represents the observed data for the
$i$th burst, and thus $p(D_i \,|\, H_x)$ is the probability of
observing $D_i$ if hypothesis $H_x$ is true. In general, we set the
priors ratio to 1, and therefore the odds ratio is the likelihood
ratio. 

The odds ratio $O_{\rm hg,bg}$ tests whether the host galaxies
predicted by a particular model are present.  We can compare different
models by forming odds ratios which compare these models; these
odds ratios would be the ratios of $O_{\rm hg,bg}$ evaluated for each
model. Equivalently, we evaluate $O_{\rm hg,bg}$ for each model, and
then compare the resulting values.  We want not only the best model,
but a model for which the host galaxies are clearly present (which
requires $O_{\rm hg,bg}>1$).  Here the models are defined by the 
value of the total burst energy, and therefore our primary objective
is an exercise in parameter estimation.  Typically for parameter estimation 
we maximize the likelihood 
for the desired parameter weighted by the prior for that parameter.
The likelihood is the numerator of the Bayes factor, i.e., 
$\prod p(D_i \,|\, H_{\rm hg})$.  If we use a uniform prior for the
total burst energy (i.e., we assume that any value of the energy is
equally probable {\it a priori}), then this likelihood is proportional
to the odds ratio (eq.~1).  Therefore maximizing the odds ratio
will give the best estimate of the total energy.  By using the odds
ratio we also demonstrate that the host galaxy model with this best
estimate of the total energy is acceptable.

For this analysis there are two types of bursts.  First are the bursts
which are localized by their gamma-ray emission (e.g., by an
Interplanetary Network or the {\it Beppo-SAX} WFC), or their X-ray
afterglow (e.g., by the {\it Beppo-SAX} NFI).  The error boxes are
dominated by the localization uncertainty and range in size from a
fraction to tens of square arcminutes; these are the error boxes which
traditionally have been searched for host galaxies. The second
category consists of the bursts followed by OTs for
which the burst positions are presumably known to a fraction of an
arcsecond.  For these bursts the localization uncertainty is small,
and the region of the sky permitted by the cosmological model may
dominate the error box.  This study shows that the bursts of the first
group place firm lower limits on the burst intensity while the second
group selects a favored range of burst intensities.  Ultimately the
observations of the second burst group will be the most constraining,
yet we will continue to include the first group for completeness and
consistency. 

The overall likelihood ratio is the product of the likelihood ratios
for each burst.  Assume that a given error box is observed down to a
limiting flux $f_{\rm lim}(\Omega)$, where we can allow this limit to vary
over the error box; $\Omega$ represents the spatial coordinates. These
observations detect $n_d$ galaxies, each with a flux $f_i$ located at
$\Omega_i$. Let the distribution of background galaxies be $\phi(f)$
(number per flux per angular area) and the burster's host galaxy is
drawn from the model-dependent distribution $\Psi(f)$, which must be
normalized to 1 (when integrated over the flux) since there can only
be one host galaxy per error box.  The burst localization uncertainty
and the host galaxy model result in a probability density
$\rho(\Omega)$ for the host galaxy's position on the sky; $\rho$ is
also normalized to 1.  Both $\Psi$ and $\rho$ represent the
cosmological model being tested. 

The likelihood ratio for one error box is 
\begin{equation}
{{p(D_i \, | \, H_{\rm hg})} \over {p(D_i \, | \, H_{\rm bg}) }} =
   \int d\Omega \int_0^{f_{\rm lim}(\Omega)} df \,\Psi(f)\rho(\Omega) + 
   \sum_{j=1}^{n_d} {{\Psi(f_j)\rho(\Omega_j)}\over {\phi(f_j) }} \quad .
\end{equation}
This expression was calculated by breaking the three dimensional space
of $f$ and $\Omega$ into little bins, evaluating the probabilities of
obtaining the observed data (galaxies in a few bins and no galaxies in
all the other bins), and then letting the bin dimensions go to zero. 

The likelihood ratio in eq.~(2) assumes the redshifts of the detected
galaxies are unknown.  When the redshift is known then both $\Psi$ and
$\phi$ in the last term in eq.~(2) gain a redshift dependence.  Of
course, some models (e.g., bursts where the intensity is a standard
candle) may give a value of $\Psi=0$ for a particular redshift.  
Redshift information will be considered in a future study.
\subsection{Data}
This analysis requires various observed distributions in a variety of
different optical bands.  Here we summarize our data sources. 

The background galaxy distribution $\phi$ is derived from galaxy
counts.  We parameterized the b$_{\rm j}$, R and K distributions using
Figure~2 of Koo \& Kron (1992) which summarizes the observations from
a number of studies.  The b$_{\rm j}$ and R distributions agree with
the study of Jones et al. (1991) while the R band distribution from
Smail et al. (1995) is a bit higher than the Koo \& Kron (1992)
distribution.  The V and I band distributions are from Smail et al.
(1995), and the U band from Jones et al. (1991).  In all cases we
extended the galaxy distribution as a power law beyond the data
presented in these sources. 

The host galaxy distribution $\Psi(f)$ is model-dependent.  This model
consists of two components:  the distribution of likely redshifts for
a given burst, and the distribution of host galaxy brightnesses at a
given redshift.  In this study we assume bursts are standard candles
whose brightness does not evolve, resulting in a unique mapping
between the burst intensity and its redshift.  In future studies we
will consider bursts with luminosity functions which evolve in time; a
luminosity function with a finite width gives a burst a range of
possible redshifts.  The host galaxy distribution at a given redshift
is also model-dependent:  the burst rate may be constant per galaxy
(e.g., Brainerd 1994) or may be proportional to the galaxy mass (e.g.,
Fenimore et al. 1993).  In many of these models the host galaxy
distribution is the regular galaxy distribution weighted by a power of
the luminosity.  Here we will assume that the burst rate is
proportional to a galaxy's luminosity, and therefore we weight the
galaxy distribution by the luminosity.  We approximate the regular
galaxy distribution by a Schechter function (Peebles 1993, p.~120), 
\begin{equation}
\psi(y) = \psi_0 y^\alpha e^{-y} \quad ,
\end{equation}
where $y=L/L_*=f/f_*$.  The intensity scale $L_*$ is
typically measured as the absolute magnitude in a given spectral band.
As described in Paper~I, we use $M_*=-19.72$ from Ratcliffe et al.
(1997) for the b$_{\rm j}$ band, $M_*=-23.12$ from Gardner et al.
(1997) for the K-band, and $M_*=-20.29$ from Lin et al. (1996)
for the R-band.  The index $\alpha$ is usually of order $-$1, and for 
computational ease we use $\alpha=-1$.
We used standard galaxy colors to interpolate the values of $M_*$ to
other optical bands. Since $M_*$ is derived from observations of
magnitude vs. redshift, to all these expressions for $M_*$ should be
added an additional term $5\log h$, where $h=H_0$/(100 km s$^{-1}$
Mpc$^{-1}$), resulting from the uncertainty in Hubble's Constant
$H_0$; however, this dependence on the value of $H_0$ is cancelled by
the $H_0$ dependence in the relationship between $z$ and the host 
galaxy flux, and therefore we do not include the dependence on $h$.  
Care must be taken that the same normalizing value of $H_0$ was used 
throughout.  In calculating the observed flux for galaxies with 
redshifts of more than a few tenths we need both $k$-corrections for the 
shift in spectrum and $e$-corrections for the evolution of the 
galaxy's luminosity and colors.  Therefore
\begin{equation}
m_* = M_* +5\log[3\times10^8 z \xi(z;q_0)]+K(z)+E(z) \ , \qquad
f_* (z) = f_0 10^{-0.4 m_*}  \ ,  
\end{equation}
where $f_0$ is the normalizing flux (i.e., the flux of a 0 magnitude
object) for a given band, and $K(z)$ and $E(z)$ are the appropriate
$k$ and $e$-corrections. This expression
assumes that $M_*$ was provided for $h=100$. The dependence on
$q_0={1\over2}\Omega_0-\Lambda_0$ is $\xi(z;q_0) = 1/q_0 +
(q_0-1)(\sqrt{1+2q_0 z}-1)/zq_0^2$ (Mattig 1958). 

We use the $k$- and $e$-corrections of Fioc \& Rocca-Volmerange (1997)
provided in the compendium of Leitherer et al. (1996).  These
corrections are given for a large number of filters by galaxy type as
a function of redshift for 3 different cosmologies---($H_0$, $\Omega_0$
and $\Lambda_0$)=(50, 0.1, 0.0), (50, 1.0, 0.0) and (75, 0.1, 0.9); in
our calculations we use the first cosmology.  We use a galaxy mix
based on Ellis (1983) to calculate a $k$- and $e$-correction for an
average $L_*$ galaxy.  Using a host galaxy model which is a weighted
average of the Schechter functions for each galaxy type would be more
accurate than using a Schechter function based on an $L_*$ with average
$k$- and $e$-corrections, but as we show below, the $k$- and
$e$-corrections change the value of the odds ratio but not the burst
energy at which it peaks.
\section{RESULTS}
We apply our methodology to two observational databases.  The first is
the compendium of Schaefer et al. (1998) which describes 23 error
boxes from before 1997 (the compendium also includes 3 of the bursts
localized by {\it Beppo-SAX}, but we treat these bursts separately).
The compendium provides the multiband magnitudes of the brightest
galaxy in the error box (except for GRB~790307, for which there is
only an upper limit); since the flux is provided for only the
brightest galaxy in the error box, this flux is also used as the
detection threshold.  Except where otherwise indicated, these
magnitudes are ``corrected'' for Galactic extinction using the
Galactic latitude $\lambda$: the extinction in band $x$ is assumed to
be $A_x = C_x (\csc(\lambda)-1)$ where $C_x$ is a constant.  The sizes
of the error boxes, as well as the bursts' energy fluences, are also
taken from Schaefer et al. (1998).  We call this database the
``Schaefer Compendium.'' 

The second database consists of the recent bursts through GRB~980703
which were followed by OTs. We do not include GRB~980425 which appears
to have originated in a supernova in a nearby galaxy (Galama et al.
1998).  If this burst is indeed associated with the supernova, the
energy requirements differ radically from other bursts
(see Table~1); in addition, no other bursts have
had nearby galaxies with supernovae in their error boxes.  Therefore
we suspect that either GRB~980425 is a member of a rare burst
population, or the association with the supernova is spurious.  Thus
this database is a complete sample of bursts which are followed by
OTs. The bursts we use are listed by Table~2, which includes the
references for the observations.  Most observations are initially
reported by IAU circulars or by circulars distributed by the GRB
Coordinates Network (GCN---Barthelmy et al. 1998).  All the OTs 
were coincident with an extended or persistent
source which we take to be the host galaxy.  We assume that the error
box, the sum of the uncertainty in the position of the OT and the
model-dependent region around the galaxy in which we expect the OT,
has a radius of 1$^{\prime\prime}$.  In the future we will use more
detailed models for the distance between the burst progenitor and the
galaxy. 

In this study the standard candle is the total energy released, which
we observe as the energy fluence.  The fluences for the Schaefer
compendium are for $E>20$~keV 
while the fluences for the OT database are predominantly the BATSE
$E=25$--2000~keV fluences; in the absence of additional spectral
information, we treat both fluence types as bolometric.  Bursts are
clearly not standard candles, as is clear from the isotropic energies
calculated for GRB~970508, GRB~971214 and GRB~980703 
which differ by a factor of
$\sim40$. Therefore in this study we do not use the redshift
information (as will be discussed in a future paper, redshift
information can be incorporated into our methodology only for burst
models with luminosity functions which allow the burst to have
occurred at a range of redshifts for a given observed brightness). 
Because we use the $k$- and $e$-correction model for ($H_0$,
$\Omega_0$ and $\Lambda_0$)=(50, 0.1, 0.0), we use the same
cosmological model in calculating the total energy from the energy
fluence, although we find that varying Hubble's constant does not
alter the qualitative results. 

To reiterate, the burst model which we investigate assumes bursts
occur in galaxies at a rate proportional to the galaxies' luminosity. 
The total burst energy $E$ (provided as an isotropic value) is
constant; for a given value of $E$ the observed fluence maps into the
burst redshift.  We calculate the odds ratio $O_{\rm hg,bg}$ (which is
also the likelihood ratio) as a function of $E$.  We want: a) the values
of $E$ where $O_{\rm hg,bg}>1$, indicating the presence of the host
galaxies predicted by the model with those values of $E$; and b) the
values of $E$ which maximize $O_{\rm hg,bg}$, indicating the preferred
range of $E$. 

Figure 1a shows $O_{\rm hg,bg}$ as a function of $E$ for the Schaefer
Compendium.  The solid curve includes the $k$- and $e$-corrections,
while the dashed curve does not.  
The two curves asymptote to 1 from below.  The brightest
galaxy in all but one error box (the error box of GRB~781104 has a
bright V=15 galaxy) is consistent with the brightest background galaxy
expected for an error box of that size.  Therefore these boxes can
rule out host galaxies of a given brightness, but cannot demonstrate
the presence of host galaxy.  This does not mean that these error
boxes have no significance since they strongly exclude low $E$ values.
Figure 1b shows similar curves for the OTs.  This
database does not exclude low $E$ values as decisively, but indicates
that $E>3\times 10^{52}$ erg is preferred.  These two databases are combined
on Figure~1c, which shows that $E\sim10^{53}$ erg is preferred. 

The odds ratios are not dominated by a few error boxes, as
demonstrated by Figure~2 which shows the odds ratio by error box for
$E=10^{51}$~erg (asterisks) and $E=10^{53}$~erg (squares).  Boxes
1--23 are the Schaefer Compendium while 24--31 are the OTs.  As can be
seen, the odds ratios for the Schaefer Compendium are mostly less than
1 for $E=10^{51}$~erg, except for GRB~781104, and they are very close
to 1 for $E=10^{53}$~erg, even for GRB~781104.  The galaxy in
GRB~781104's error box is much brighter than $L_*$ for the distance to
the burst expected for $E=10^{53}$~erg, and it falls far out on the
Schechter function's exponential; this galaxy is therefore unlikely to
be the host for this value of $E$.  On the other hand, the galaxies
associated with the OTs are much fainter than the host galaxies
expected for $E=10^{51}$~erg, and thus are more likely to be
background galaxies; therefore the odds ratios for these boxes are
less than 1.  However, for $E=10^{53}$~erg these observed galaxies are
consistent with the predicted host galaxies, and the odds ratios are
greater than 1. 

Figure 3a shows the $O_{\rm hg,bg}$ curves vs. $E$ for $H_0=65$ km
s$^{-1}$ Mpc$^{-1}$ instead of $H_0=50$ km s$^{-1}$ Mpc$^{-1}$.  Note 
that the $k$- and $e$-corrections still assume $H_0=50$ km s$^{-1}$ 
Mpc$^{-1}$.  As can be seen, this figure barely differs from Figure~1b.  
On the other hand, Figure~3b shows the same curves if we assume the 
radius of the error box (in this case the distance between the burst 
and the galaxy) is 0.5$^{\prime\prime}$ instead of 1$^{\prime\prime}$. 
In this case the odds ratios are shifted up significantly because the 
probability that the observed galaxy is an unrelated background galaxy 
has decreased in proportion to the square of the radius (i.e., the area of 
the error box) for {\it each} error box.  Nonetheless, the same $E$ 
range is preferred.

Figure~4 shows the effect of changing the value of $M_*$ by $\pm1$.
Increasing $M_*$ means we expect the galaxies to be fainter at a given
distance, and therefore the host galaxies can be closer and the bursts can
be intrinsically fainter; the opposite is expected if $M_*$ decreases.
As can be seen, changing $M_*$ by 1 shifts the energy at which the
odds ratio peaks by less than a factor of 2.
\section{DISCUSSION}
There are now both theoretical and observational arguments that bursts
are further and more energetic than predicted by the minimal
cosmological model.  Theoretically, the source models where the
progenitor is a rare endpoint of stellar evolution lead to source
evolution models where the burst rate is proportional to the star
formation rate (Totani 1997; Wijers et al. 1998; Hartmann \& Band
1998).  The evolution in the source density balances the cosmological
curvature of space, and the intensity distribution is consistent with
more distant bursts, although quantitative discrepancies need to be 
resolved (Petrosian \& Lloyd 1998; Hartmann \& Band 1998).

The three bursts with redshifts---GRB~970508 at $z=0.835$ (Metzger et 
al. 1997; Bloom et al. 1998), GRB~980703 at $z=0.966$ (Djorgovski
et al. 1998a) and GRB~971214 at $z=3.4$ (Kulkarni et 
al. 1998)---are further than predicted by the minimal model for their 
intensities.  But currently there are only three redshifts.  Similarly, 
the host galaxies (or upper limits) for the OTs are 
fainter than expected for the minimal model.  Here we have quantified 
this perception that the host galaxies are faint, and derived the 
implied standard candle total energy.

However, the burst energy is not a constant for all bursts, as
demonstrated by Table~1, and therefore bursts must be
characterized by luminosity functions, as we will investigate in a
future paper.  Nonetheless, our results show that on average the burst
energy is significantly greater than previously thought.  The
theoretical consequences are already being studied. 
\section{SUMMARY}
In Paper~I we developed a methodology to determine whether a host
galaxy predicted by a specified model is present within a burst error
box.  This methodology is also applicable to bursts whose positions
are known with negligible uncertainty (e.g., bursts followed by OTs)
because the relevant error box is the sum of the positional
uncertainty and the model-dependent region around the host galaxy in
which the burst could have occurred.  In Paper~I we verified the
absence of the host galaxies predicted by the ``minimal'' model where
bursts do not undergo density or luminosity evolution.  Here we
applied this methodology to two databases, the first a set of 23
moderate-sized error boxes from before 1997, and the second the recent
bursts followed by OTs. We used a burst model where bursts occur
within 1$^{\prime\prime}$ of the host galaxy and have the same
standard candle total energy.  We allowed the total burst energy to
vary, and found the energy range consistent with the galaxies in the
error boxes.  To satisfy the observed intensity distribution, the
source density must have evolved, as has indeed been suggested. 

We found that the pre-1997 error boxes strongly rule out isotropic
burst energies below $10^{52.5}$ erg, while the OTs favor energies of
$\sim10^{53}$ erg.  This result is relatively insensitive to the
value of Hubble's constant and the $k$- and $e$-corrections. 

In a future study we will consider burst models with luminosity 
functions.  Eventually our host galaxy methodology will be combined 
with analyses of other data (e.g., the burst intensity distribution) 
to develop a burst model consistent with all observations.
\acknowledgments
D.~Band's gamma-ray
burst research is supported by the {\it CGRO} guest investigator
program and NASA contract NAS8-36081.  D.~Hartmann acknowledges
support from the {\it CGRO} guest investigator program. 

\clearpage

\figcaption{The odds ratio $O_{\rm hg,bg}$ as a function of the 
standard candle burst energy $E$ (assumed to have been radiated 
isotropically).  The solid curve includes $k$- and $e$-corrections 
whereas the dashed curve does not.  The assumed cosmological model is 
$H_0=50$ km s$^{-1}$ Mpc$^{-1}$, $\Omega_0=0.1$ and $\Lambda_0=0$.  Panel 
1a uses the pre-1997 bursts from Schaefer et al. (1998), 1b 
uses the recent 
bursts followed by optical transients, and 1c uses both 
databases.  $O_{\rm hg,bg}\ll 1$ indicates the absence of the host 
galaxy predicted by the model with the given value of $E$, while a 
maximum value of $O_{\rm hg,bg}$ shows the most likely value of $E$.}
\figcaption{Distribution of the odds ratio $O_{\rm hg,bg}$ by burst 
for $E=10^{51}$ erg (asterisks) and $E=10^{53}$ erg (squares).  Bursts 
1--23 are the pre-1997 bursts from Schaefer et al. (1998):
1.~GRB~781104; 2.~GRB~781119; 3.~GRB~781124; 4.~GRB~790113; 
5.~GRB~790307; 6.~GRB~790313; 7.~GRB~790325; 8.~GRB~790329; 
9.~GRB~790331; 10.~GRB~790406; 11.~GRB~790418; 12.~GRB~790613; 
13.~GRB~791105; 14.~GRB~791116; 15.~GRB~910122; 16.~GRB~910219; 
17.~GRB~911118; 18.~GRB~920325; 19.~GRB~920406; 20.~GRB~920501; 
21.~GRB~920711; 22.~GRB~920720; and 23.~GRB~920723.
Bursts 24--31 are the recent bursts followed by an optical transient:
24.~GRB~970228; 25.~GRB~970508; 26.~GRB~971214; 27.~GRB~980326; 
28.~GRB~980329; 29.~GRB~980519; 30.~GRB~980613; and 31.~GRB~980703.}
\figcaption{The same as figure 1b except in panel 3a a value of 
$H_0=65$ km s$^{-1}$ Mpc$^{-1}$ is used, while in panel 3b the radius of 
the error box surrounding the burst is decreased by a factor of 2.} 
\figcaption{The dependence of the odds ratio on the value of $M_*$.
$M_*$ has been increased (dashed curve) or decreased (dot-dashed curve)
by 1 compared to the currently accepted value (solid curve).  The
calculation assumes $H_0=50$ km s$^{-1}$ Mpc$^{-1}$, $\Omega_0=0.1$ and 
$\Lambda_0=0$, and the $k-$ and $e-$corrections are included.}

\clearpage

\begin{deluxetable}{l c c c c c}
\tablecolumns{6}
\scriptsize
\tablewidth{0pc}
\tablecaption{Energies of Bursts with Redshifts}
\tablehead{
\colhead{Burst} & 
\colhead{$z$} & 
\colhead{Ref.} &
\colhead{Fluence\tablenotemark{a}} & 
\colhead{Peak Flux\tablenotemark{b}} & 
\colhead{Energy \tablenotemark{c}} 
}
\startdata
% $\times10^{}$ 
GRB~970508 &  0.835               & d & $3.96\times10^{-6}$ & 0.97 & 
$6.50\times10^{51}$ \nl
GRB~971214 &  3.42                & e & $1.09\times10^{-5}$ & 1.95 &
$2.95\times10^{53}$ \nl
GRB~980425 & $8.43\times 10^{-3}$ & f & $4\times10^{-6}$    & 0.96 &
$7.24\times10^{47}$ \nl
GRB~980703 & 0.966                & g & $4.59\times10^{-5}$ & 2.42 &
$1.03\times10^{53}$ \nl
\enddata
\tablenotetext{a}{Fluence greater than 25~keV, erg cm$^{-2}$, assumed to
be bolometric.  From the BATSE catalog---Meegan et al. (1998).}
\tablenotetext{b}{Peak photon flux in the 50--300~keV band accumulated
over 1.024~s.  From the BATSE catalog---Meegan et al. (1998).}
\tablenotetext{c}{Total burst energy if radiated isotropically.  Assumes
$H_0=65$ km s$^{-1}$ Mpc$^{-1}$, $\Omega=0.3$, and $\Lambda=0$.}
\tablenotetext{d}{Metzger et al. (1997); Bloom et al. (1998).}
\tablenotetext{e}{Kulkarni et al. (1998).}
\tablenotetext{f}{Galama et al. (1998).}
\tablenotetext{g}{Djorgovski et al. (1998a).}
\end{deluxetable}
\clearpage

\begin{deluxetable}{l c c c c c}
\tablecolumns{6}
\scriptsize
\tablewidth{0pc}
\tablecaption{The Host Galaxies Associated with Optical Transients}
\tablehead{
\colhead{Burst} & 
\colhead{Fluence\tablenotemark{a}} & 
\colhead{$R_{\rm det}$\tablenotemark{b}} & 
\colhead{Ref.} & 
\colhead{Ext.\tablenotemark{c}} &
\colhead{$R_{\rm corr}$}
}
\startdata
GRB~970228 & $4.6\times10^{-6}$\tablenotemark{d} & 25.2 & e & 0.65 & 24.6 \nl
GRB~970508 & $3.96\times10^{-6}$ & 25.72 & f & 0.17 & 25.55 \nl
GRB~971214 & $1.09\times10^{-5}$ & 25.6 & g & 0.01 & 25.6 \nl
GRB~980326 & $1\times10^{-6}$ & 25.5 & h & 0.20 & 25.3 \nl
GRB~980329 & $8.26\times10^{-5}$ & 25.7 & i & 0.31 & 25.4 \nl
GRB~980519 & $2.54\times10^{-5}$ & 25.55 & j & 0.85 & 24.7 \nl
GRB~980613 & $1.71\times10^{-6}$\tablenotemark{k} & 24.5 & l & 0.07 & 24.4 \nl
GRB~980703 & $4.59\times10^{-5}$ & 22.3 & m & 0.14\tablenotemark{m} & 22.2 \nl
\enddata
\tablenotetext{a}{Fluence greater than 25~keV, erg cm$^{-2}$, from the
BATSE catalog (Meegan et al. 1998), unless otherwise indicated.}
\tablenotetext{b}{$R$ magnitude of detected galaxy.}
\tablenotetext{c}{Extinction from Burstein \& Heiles (1982) quoted by
Hogg \& Fruchter (1998), unless otherwise indicated.}
\tablenotetext{d}{Palmer et al. (1998).}
\tablenotetext{e}{HST observation of extended source reported by
Fruchter et al. (1998).}  
\tablenotetext{f}{Galaxy at $z=0.835$ observed by Bloom et al.
(1998).} 
\tablenotetext{g}{Extended source observed by Kulkarni et al. (1998)
with $z=3.418$.} 
\tablenotetext{h}{Galaxy observed by Djorgovski et al. (1998b), GCN~57.}
\tablenotetext{i}{Galaxy observed by Djorgovski et al. (1998c), GCN~41.}
\tablenotetext{j}{H. Pedersen quoted by Hogg \& Fruchter (1998).}
\tablenotetext{k}{Woods et al. (1998), GCN~112.}
\tablenotetext{l}{Djorgovski et al. (1998d), GCN~117.}
\tablenotetext{m}{Djorgovski et al. (1998a).}
%
%\tablenotetext{}{Galaxies in field of X-ray transient, Djorgovski et al.
%(1998), GCN~25.}
%
\end{deluxetable}

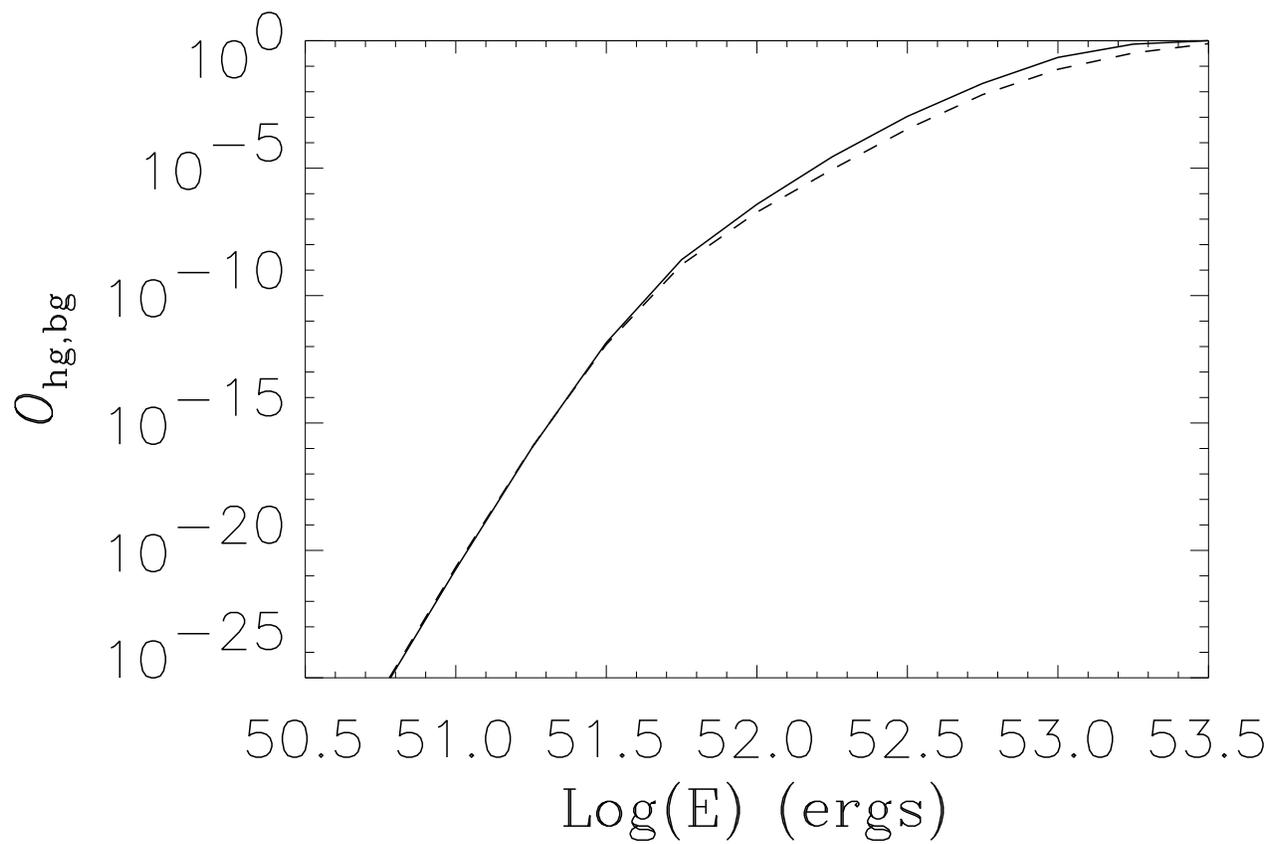
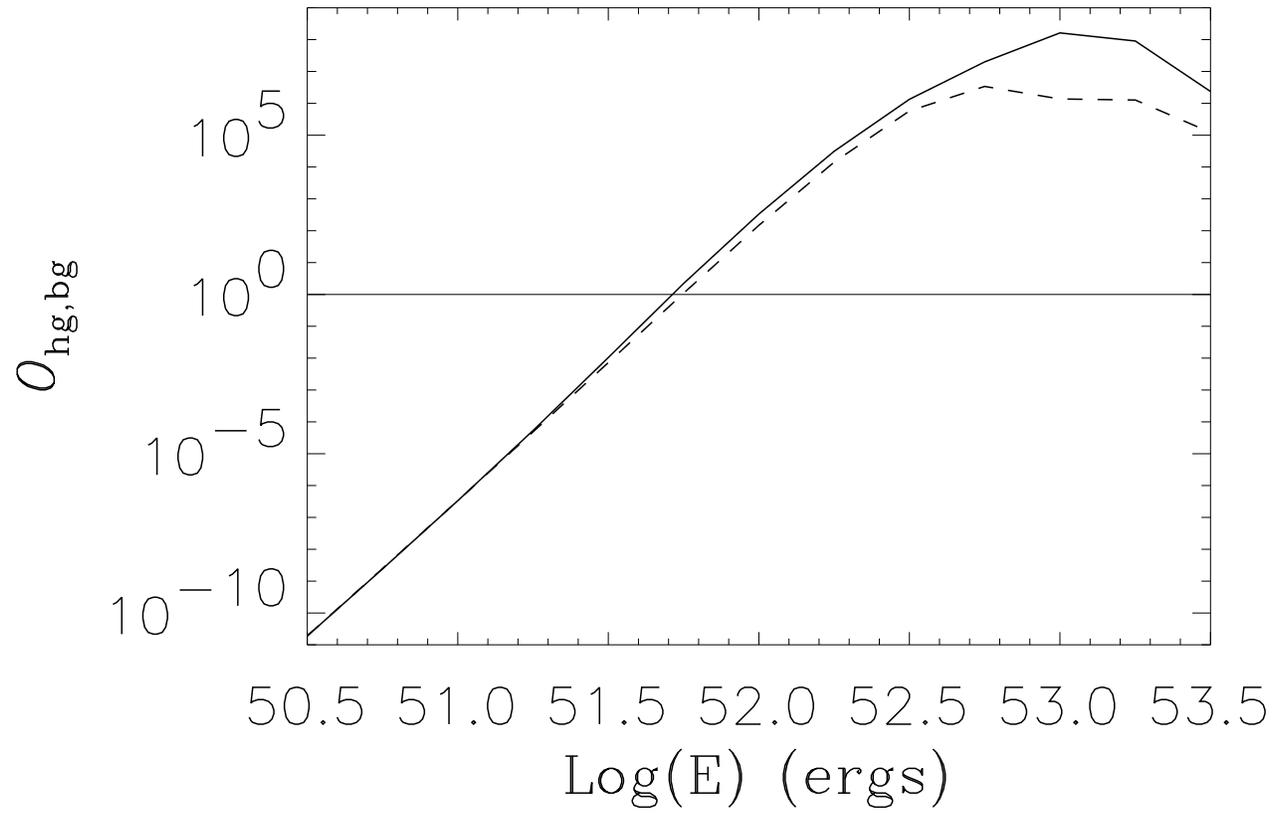
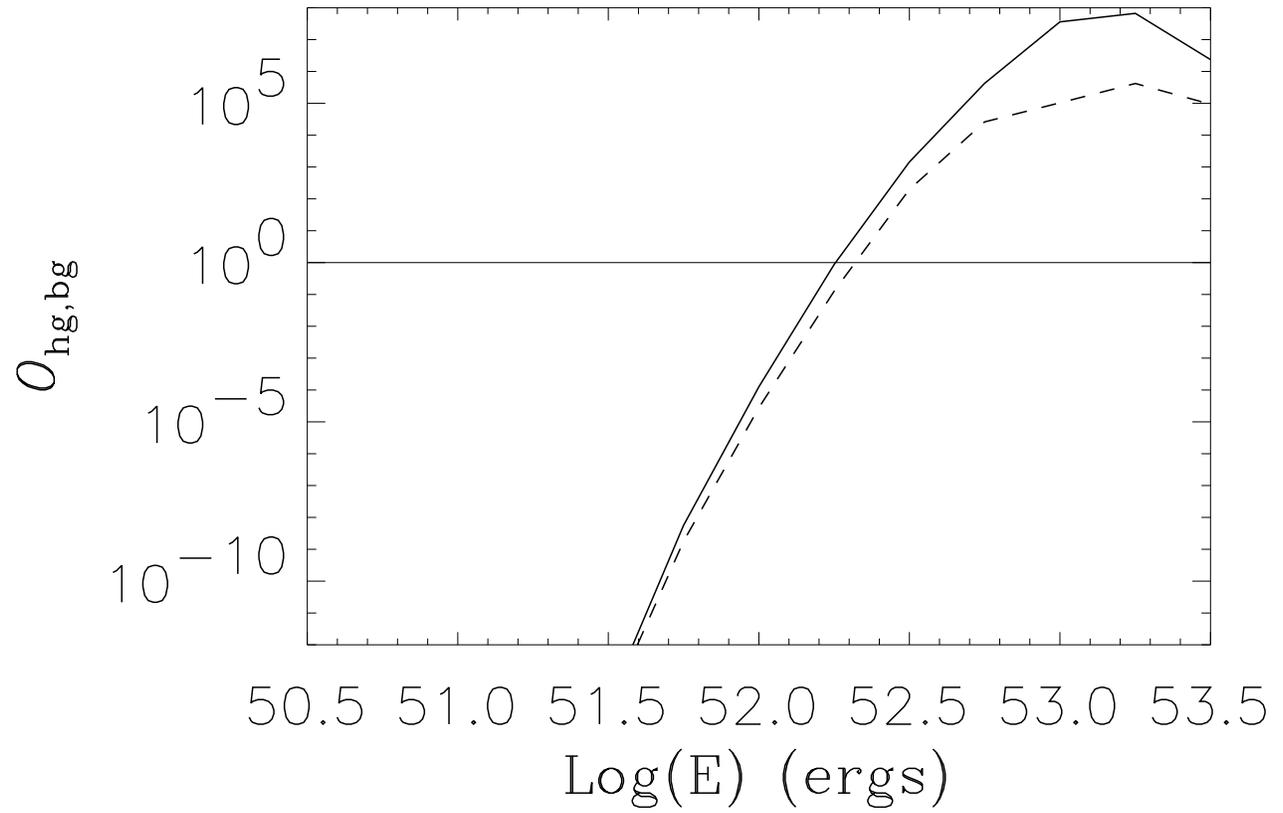
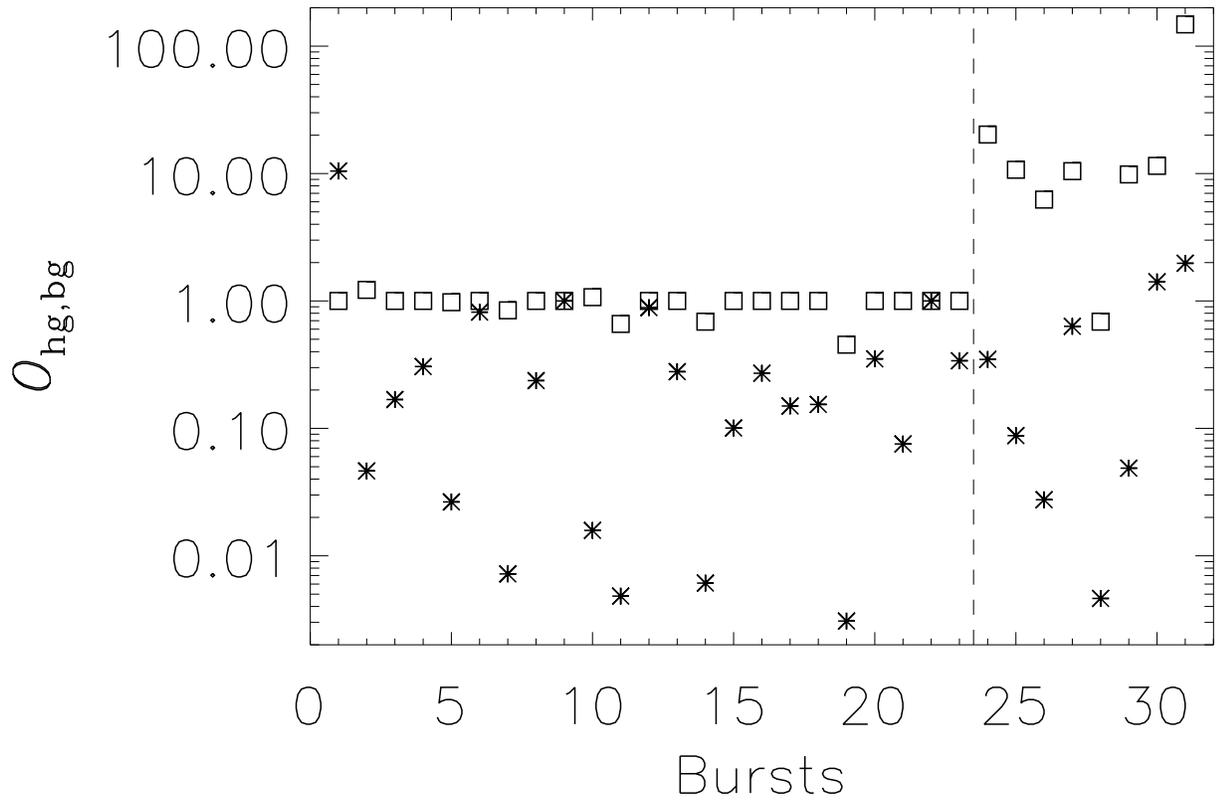
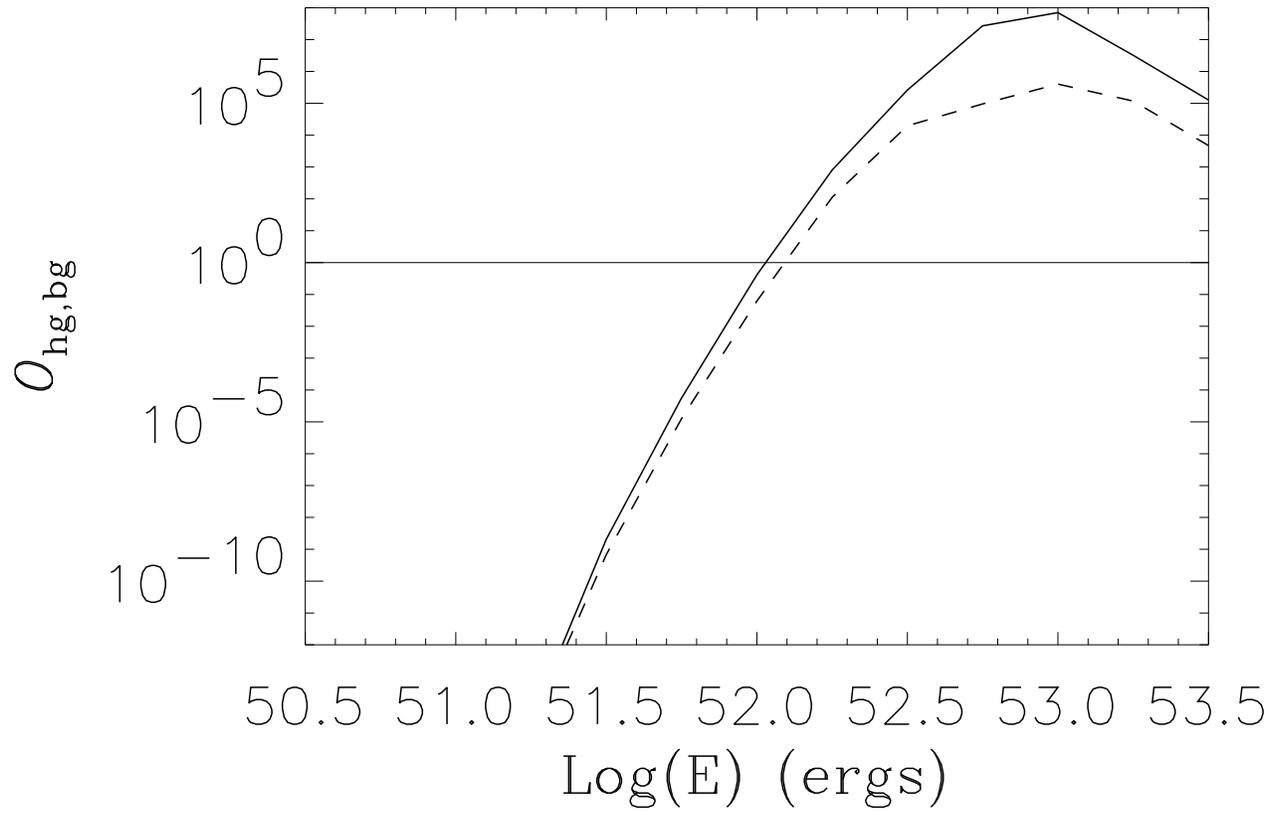
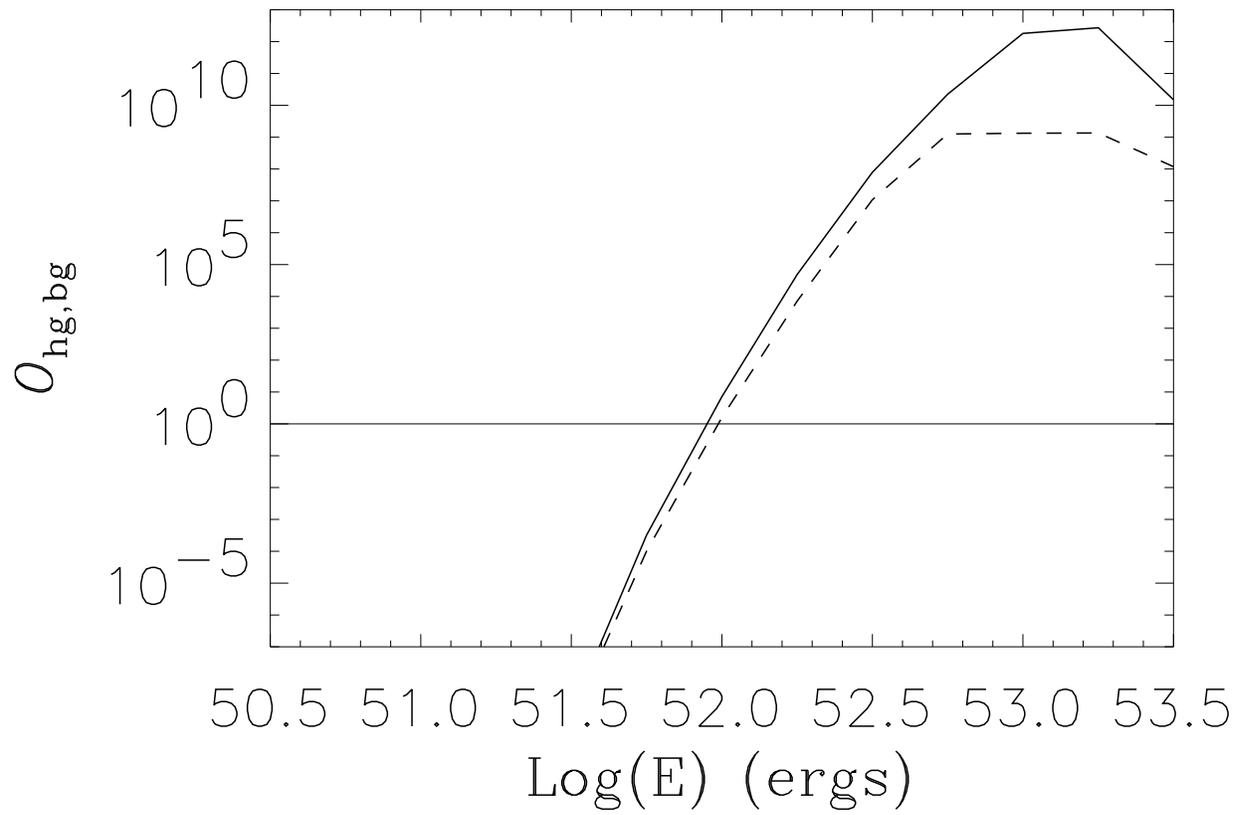
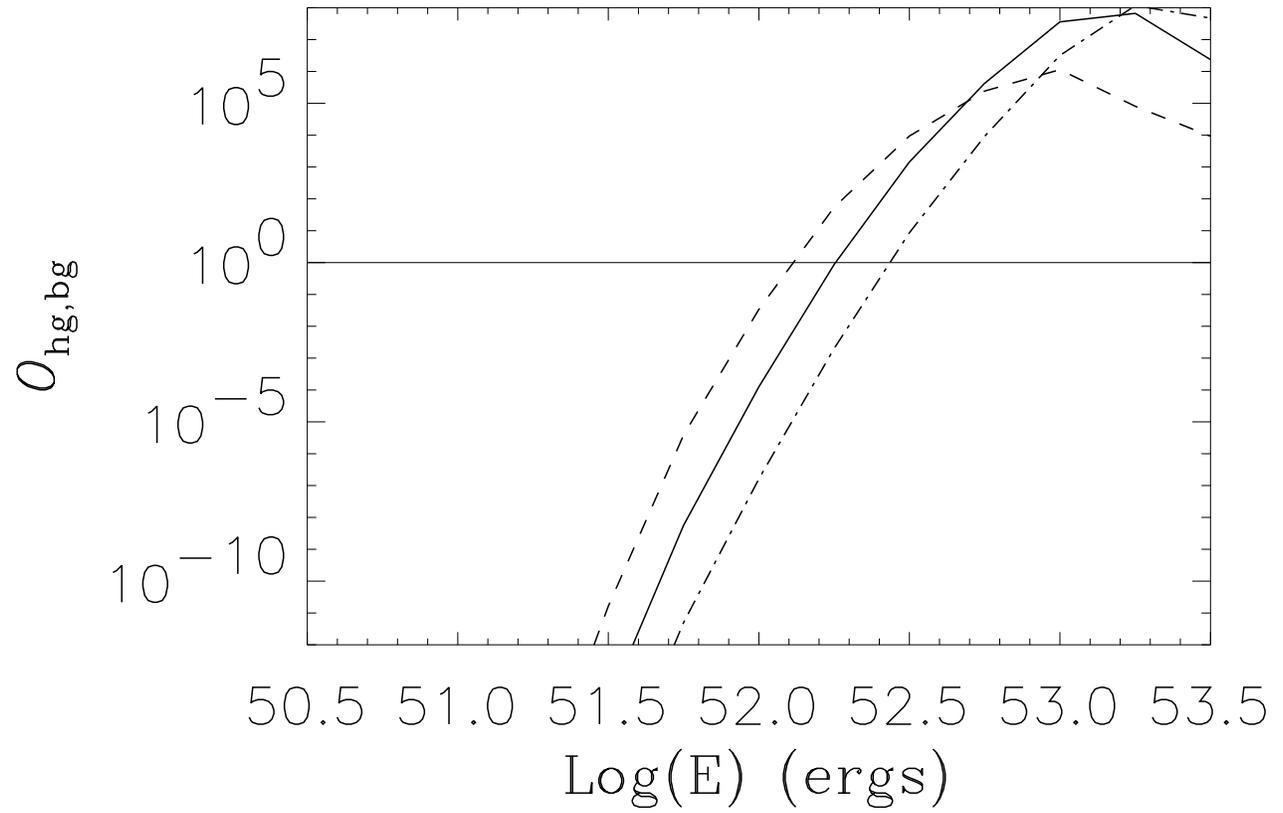
\end{document}